\documentclass[journal,twoside,letterpaper]{IEEEtranTCOM}

\usepackage{graphicx}
\usepackage{amssymb}
\usepackage{epstopdf}
\usepackage[cmex10]{amsmath}
\usepackage{amsthm}
\usepackage{enumerate}
\usepackage{eufrak}
\usepackage{cite}
\usepackage{mathcomp}
\usepackage{supertabular}
\usepackage{longtable}
\usepackage{stmaryrd}
\usepackage{url}
\usepackage{color}
\usepackage{rotating}
\usepackage{float}
\usepackage{tikz}
\usetikzlibrary{shapes}
\usepackage[boxed]{algorithm2e}
\usepackage{array}
\usepackage{placeins}
\usepackage{multicol}

\renewcommand{\theenumi}{\roman{enumi}}

\renewcommand{\theenumii}{\alph{enumii}}

\newcolumntype{C}[1]{>{\centering\let\newline\\\arraybackslash\hspace{0pt}}m{#1}}

\theoremstyle{definition}
\renewcommand{\thesection}{\arabic{section}}

\newtheorem{prop}{Proposition}[section]
\newtheorem{thm}{Theorem}[section]
\newtheorem{cor}{Corollary}[section]
\newtheorem{lem}{Lemma}[section]

\newtheorem{defn}{Definition}[section]
\newtheorem{exa}{Example}[section]

\newtheorem{rem}{Remark}

\theoremstyle{remark}

\renewcommand\Sigma{\mathcal{X}}
\renewcommand\Gamma{\mathcal{Y}}

\begin{document}

\title{Importance of Symbol Equity in Coded Modulation for Power Line Communications}
\author{Yeow Meng Chee,~\IEEEmembership{Senior Member, IEEE}, Han Mao Kiah, 
\\ Punarbasu Purkayastha,~\IEEEmembership{Member, IEEE}, 
and Chengmin Wang
\thanks{Research of Y.~M.~Chee, H.~M.~Kiah, P.~Purkayastha, 
and C.~Wang is supported in part by the National Research Foundation of Singapore under Research Grant NRF-CRP2-2007-03. 
C.~Wang is also supported in part by NSFC under Grant No. 11271280.}
\thanks{Y.~M.~Chee, H.~M.~Kiah and P.~Purkayastha 
are with the Division~of~Mathematical Sciences,
  School~of~Physical~and~Mathematical~Sciences,
  Nanyang~Technological~University, 21~Nanyang~Link, Singapore~637371,
  Singapore (emails:\{ymchee, hmkiah, punarbasu\}@ntu.edu.sg).}%
\thanks{C.~Wang is with School~of~Science, Jiangnan~University, Wuxi 214122, China (email:wcm@jiangnan.edu.cn).}
\thanks{This paper was presented in part at the 2012 IEEE International Symposium on Information Theory \cite{Cheeetal:2012a}.}
}


\newcommand{\vA}{{\sf A}}
\newcommand{\vB}{{\sf B}}
\newcommand{\vC}{{\sf C}}
\newcommand{\vc}{{\sf c}}
\newcommand{\vu}{{\sf u}}
\newcommand{\vv}{{\sf v}}
\newcommand{\vw}{{\sf w}}
\newcommand{\tA}{\textrm A}
\newcommand{\tB}{\textrm B}
\newcommand{\A}{\mathcal A}
\newcommand{\B}{\mathcal B}
\newcommand{\C}{\mathcal C}
\newcommand{\D}{\mathcal D}
\newcommand{\G}{\mathcal G}
\newcommand{\R}{\mathcal R}
\newcommand{\SSS}{\mathcal S}

\newcommand{\enarrow}{e_{\sf N}}
\newcommand{\efade}{e_{\sf F}}
\newcommand{\eimpulse}{e_{\sf IMP}}
\newcommand{\einsert}{e_{\sf INS}}
\newcommand{\edelete}{e_{\sf DEL}}
\newcommand{\ec}{E(\mathcal C)}
\newcommand{\Ln}{L_{\sf N}}
\newcommand{\Lf}{L_{\sf F}}
\newcommand{\pnarrow}{P_{\sf NBD}}

\newcommand{\union}{\bigcup\limits}
\newcommand{\swt}{\textrm {swt}}

\newcommand{\CC}{\mathbb C} 
\newcommand{\RR}{\mathbb R}
\newcommand{\ZZ}{\mathbb Z}
\newcommand{\FF}{\mathbb F}
\newcommand{\ceiling}[1]{\left\lceil{#1}\right\rceil}
\newcommand{\floor}[1]{\left\lfloor{#1}\right\rfloor}

\newcommand{\wt}[1]{\textrm{wt}{(#1)}}
\newcommand{\trace}[1]{\textrm{Trace}{(#1)}}
\newcommand{\supp}[1]{\textsf{supp}{(#1)}}
\newcommand{\lev}[1]{\textsf{lev}{(#1)}}
\newcommand{\dist}{\textsf{dist}}
\newcommand{\packing}[1]{\textrm{Packing}_\textrm{#1}}
\newcommand{\ppty}[1]{\textsf{Property #1}}
\newcommand{\pp}{^\prime}

\newcommand{\block}{\mathcal B}
\newcommand{\gbtd}{\textrm{GBTD}}
\newcommand{\bibd}{\textrm{BIBD}}
\newcommand{\rbibd}{\textrm{RBIBD}}
\newcommand{\pbd}{\textrm{PBD}}
\newcommand{\drtd}{\textrm{DRTD}}
\newcommand{\kts}{\textrm{KTS}}
\newcommand{\fkts}{\textrm{FKTS}}
\newcommand{\td}{\textrm{TD}}
\newcommand{\ttd}{\textrm{TTD}}

\newcommand{\inprod}[1]{\langle{#1}\rangle}


\newcommand{\beas}{\begin{eqnarray*}} 
\newcommand{\eeas}{\end{eqnarray*}} 

\newcommand{\bm}[1]{{\mbox{\boldmath $#1$}}} 

\newcommand{\tworow}[2]{\genfrac{}{}{0pt}{}{#1}{#2}}
\newcommand{\qbinom}[2]{\left[ {#1}\atop{#2}\right]_q}

\newcommand{\Lovasz}{Lov\'{a}sz }
\newcommand{\citereq}{[citation required]}
\newcommand{\etal}{\emph{et al.}}

\maketitle

\begin{abstract}
The use of multiple frequency shift keying modulation with permutation codes addresses the problem of 
permanent narrowband noise disturbance in a power line communications system.
In this paper, we extend this coded modulation scheme based on permutation codes to general codes and introduce an
additional new parameter that more precisely captures a code's performance against permanent narrowband noise.
As a result, we define a new class of codes, namely, equitable symbol weight codes, which are optimal with respect to this measure.
\end{abstract}

\begin{IEEEkeywords}
multiple frequency shift key modulation, power line communications, narrowband noise, 
equitable symbol weight codes
\end{IEEEkeywords}


\section{Introduction}
Power line communications (PLC) is a 
technology that enables the transmission of data over 
electric power lines.
It was started in the 1910's for voice communication \cite{Schwartz:2009}, and used in
the 1950's in the form of ripple control for load and tariff management
in power distribution.
With the emergence of the Internet in the 1990's, research into broadband PLC
gathered pace as a promising technology for Internet access and local area networking, since
the electrical grid infrastructure provides ``last mile'' connectivity
to premises and capillarity within premises. Recently, there has been a renewed interest
in high-speed narrowband PLC due to applications in sustainable energy strategies, specifically in
smart grids (see \cite{Haidineetal:2011,Dzungetal:2011,Liuetal:2011,ZhangYang:2011}).

However, power lines present a difficult communications environment and
overcoming permanent narrowband disturbance has remained a 
challenging problem \cite{Vinck:2000,Biglieri:2003,Pavlidouetal:2003}. 
Vinck \cite{Vinck:2000} addressed this problem by 
showing that multiple frequency shift keying (MFSK) modulation, 
in conjunction with the use of a 
permutation code having minimum (Hamming) distance $d$,
is able to correct up to $d-1$ errors due to narrowband noise.
Since then, more general codes such as 
constant-composition codes (see \cite{Luoetal:2003,DingYin:2005a,DingYin:2005b,DingYuan:2005,Chuetal:2006,DingYin:2006,Cheeetal:2007,Cheeetal:2010a,Huczynska:2010,GaoGe:2011}),
frequency permutation arrays (see \cite{HuczynskaMullen:2006,Huczynska:2010}), 
and injection codes (see
\cite{Dukes:2011})
have been considered as possible replacements for permutation codes in PLC.
Versfeld \etal \cite{Versfeldetal:2005,Versfeldetal:2010}
later introduced the notion of  `same-symbol 
weight' (henceforth, termed as {\em symbol weight}) of a 
code as a measure of the capability of a code in dealing with narrowband noise.
They also showed empirically that
low symbol weight cosets of Reed-Solomon codes 
outperform normal Reed-Solomon codes
in the presence of narrowband noise and additive white Gaussian noise.
Sizes of symbol-weight spaces were investigated by Chee \etal \cite{Cheeetal:2012c} recently.

Unfortunately, symbol weight alone is not sufficient to capture the performance
of a code in dealing with permanent narrowband noise.
The purpose of the paper is to extend the analysis of
Vinck's coded modulation scheme based on permutation codes
(see \cite{Vinck:2000},\cite[Subsection 5.2.4]{Ardakanietal:2010})
to general codes. In the process, we introduce
an additional new parameter that more precisely captures a code's performance against permanent
narrowband noise. This parameter is related to {\em symbol equity},
the uniformity of frequencies of symbols in each codeword.
Codes designed taking into account this new parameter, or {\em equitable symbol weight codes},
are shown to perform better than general ones.

The current proposed standards, such as ITU-T Recommendation G.9902 (G.hnem) and IEEE P1901.2, for
communication over narrowband power line channel use Orthogonal Frequency
Division Multiplexing (OFDM) based modulation schemes instead of FSK based
schemes.
In contrast to MFSK scheme which uses only one frequency at a time, OFDM
uses multiple frequencies at the same time for transmitting information.
Preliminary results on extensions of the current work to use multiple
frequencies are presented in \cite{Cheeetal:2013}. Further investigations
and comparisons with current OFDM based schemes are an interesting avenue
for future research. Finally, we remark that the notion of symbol equity
discussed in this work is also applicable to systems where criss-cross
types of errors are encountered \cite{Plass:2008}.

The outline of the rest of the paper is as follows.
In Section 2 we introduce the basic definitions and notation.
In Section 3 we introduce the noise model and the criterion under which correct decoding can be performed. 
In particular, we introduce a new parameter that captures how well a code can perform under narrowband noise. 
In Section 4 we show that equitable symbol weight codes are optimal with respect to this new parameter. 
We present some simulation results in Section 5 to compare the performance of equitable symbol weight codes 
with other block codes previously studied in the literature. 

\section{Preliminaries}

We denote the set of integers and positive integers by  $\ZZ$ and $\ZZ_{>0}$ respectively.
We denote the set $\{1,\dots,n\}$ by the notation $[n]$. For a finite set
$X$, the collection of all subsets of $X$, or the \emph{power set} of $X$,
is denoted by $2^X.$

Let $T$ be an index set and $\Sigma$ be a set of {\em symbols}. 
We denote {\em a sequence or a vector with index set $T$} by $(u_t:t\in T,u_t\in\Sigma)$.
In contrast, we denote a {\em multiset} by angled brackets, that is, $\langle u_t: t\in T\rangle$.
For the latter, when more convenient, the exponential notation 
$\langle u_1^{t_1}u_2^{t_2}\cdots u_n^{t_n}\rangle$ is used to describe a multiset
with exactly $t_i$ elements $u_i$, $i\in [n]$.

When $|\Sigma|=q$,  a {\em $q$-ary code} 
$\mathcal C$ of {\em length} $n$ over the {\em alphabet} $\Sigma$ is a 
subset of $\Sigma^n$. Elements of $\mathcal C$ are called {\em codewords}.
The {\em size} of $\mathcal C$ is the number of codewords in $\mathcal C$. 
For $i\in[n]$, the $i$th {\em coordinate} of a 
codeword $\vu$ is denoted by $\vu_i$.

\subsection{Symbol weight}

Let $\vu\in\Sigma^n$.
For $\sigma\in \Sigma$, $w_\sigma(\vu)$ is the number of times the symbol $\sigma$
appears among the coordinates of $\vu$, that is, 
\begin{equation*}
w_\sigma(\vu)=|\{i\in[n]: \vu_i=\sigma\}|.
\end{equation*}
\noindent The {\em symbol weight} of $\vu$ is 
\begin{equation*}
\text{swt}(\vu)=\max_{\sigma\in \Sigma} w_\sigma(\vu).
\end{equation*}

A code has {\em bounded symbol weight} $r$
if the maximum symbol weight of all its codewords is $r$. 
A code $\mathcal C$ has {\em constant symbol weight} 
$r$ if all its codewords have symbol weight exactly $r$.
For any $\vu\in\Sigma^n$, observe that $\text{swt}(\vu)\geq \lceil{n/q}\rceil$.
A code has {\em minimum symbol weight}
if it has constant symbol weight $\lceil{n/q}\rceil$.

A codeword $\vu\in\Sigma^n$ is said to have {\em equitable symbol weight}
if $w_\sigma(\vu)\in \{\lfloor{n/q}\rfloor,\lceil{n/q}\rceil\}$ for
all $\sigma\in \Sigma$. 
In other words, if $r=\ceiling{n/q}$, then every symbol appears $r$ or $r-1$ times in $\vu$.
If all the codewords of $\mathcal C$ have equitable symbol weight,
then the code $\mathcal C$ is called an 
{\em equitable symbol weight code}. 
Every equitable symbol weight code is hence a minimum symbol weight code.

\subsection{Composition and Partition}

The {\em composition} of $\vu\in\Sigma^n$ 
is the sequence $(w_\sigma(\vu): {\sigma\in \Sigma})$, 
and the {\em partition} of $\vu$ is the multiset $\langle w_\sigma(\vu): \sigma\in \Sigma\rangle$.

Fix a multiset of nonnegative numbers $\langle c_\sigma: \sigma \in \Sigma\rangle$ 
such that $\sum_{\sigma\in\Sigma} c_\sigma = n$.
A code $\C$ is a {\em constant composition code with composition $(c_\sigma:\sigma\in \Sigma)$ } 
if  all words in $\C$ have composition $(c_\sigma:\sigma\in \Sigma)$. 
Similarly, a code $\C$ is a {\em constant partition code with partition $\langle c_\sigma:\sigma\in \Sigma\rangle$ } 
if all words in $\C$ have partition $\langle c_\sigma:\sigma\in \Sigma\rangle$.

Clearly, a constant composition code is necessarily a constant partition code. 
The following example demonstrates that the converse is not true.

\begin{exa}
The code $\{(1,2,3),(2,3,4),(3,4,1),(4,1,2)\}$ is a constant partition code with
partition $\langle 1^3 0\rangle$,
 since in each code word three symbols appear once each, 
 and one symbol does not appear.
However, the words have different compositions.
\end{exa}

We show that an equitable symbol weight code is necessarily a constant partition code 
with minimum symbol weight.
This follows from the next lemma that states that for any $\vu\in\Sigma^n$ having equitable symbol
weight, the number of symbols occurring with frequency $\lceil{n/q}\rceil$ in $\vu$
is uniquely determined. Hence, the frequencies of symbols in an
equitable symbol weight codeword are as uniformly distributed as possible and
the partition of the codeword is fixed.

\begin{lem}\label{lem:comp}
Let $\vu\in\Sigma^n$,
$r=\lceil{n/q}\rceil$, and $t=qr-n$.
If $\vu$ has equitable symbol weight, 
then $\vu$ has partition $\langle r^{q-t}(r-1)^t\rangle$
\end{lem}

\begin{IEEEproof}
Let $x=|\{\sigma\in \Sigma: w_\sigma(\vu)=r\}|$ and 
$y=|\{\sigma\in \Sigma: w_\sigma(\vu)=r-1\}|$. Then the following equations hold:
\begin{equation*}
x+y = q,\mbox{ and }
rx+(r-1)y= n.
\end{equation*} 
Solving this set of equations gives the lemma.
\end{IEEEproof}

Using the above notation, we observe that equitable symbol weight codes are generalizations 
of certain classes of codes which have been studied in PLC applications.
For example, if $q|n$, then an equitable symbol weight code has constant partition $\langle (n/q)^q\rangle$, 
which is known as a frequency permutation array (FPA). 
If $n \le q$ then an equitable symbol weight code has constant partition $\langle 1^n0^{q-n}\rangle$, 
which is called an injection code. 
Finally, if $n = q$, then all definitions coincide to give the definition of a permutation code.

\begin{figure}[hbtp]
\center
\scriptsize
\begin{tikzpicture}[thick,scale=0.7, every node/.style={scale=0.8}]

\draw[thick](-4.5,-5.5) rectangle +(9, 11);
\filldraw[fill=lightgray,very thick] (0,0) ellipse (3.2cm and 3.2cm) ;

\draw[thick] (0,0.7) ellipse (4.2cm and 4.2cm);
\draw[thick] (0,-0.9) ellipse (4.2cm and 4.2cm);
\draw[thick] (0,-1.5) ellipse (3cm and 3cm);
\draw[thick] (0,-1) ellipse (2cm and 2cm) ;
\draw[thick] (0,0.2) ellipse (2cm and 2cm) ;

\draw(0,4) node{Codes with minimum symbol weight};
\draw (0,-4.7)node{Constant Partition Codes};
\draw(0,2.3) node {\bf Equitable Symbol Weight Codes};	
\draw(0,-3.7) node {Constant Composition Codes};
\draw(0,-2.5) node {FPAs};
\draw(0,1.7) node {Injection Codes};
\draw(0,0) node {\bf Permutation Codes};
\end{tikzpicture}
\caption{Generalizations of Permutation Codes}
\label{fig:plccodes}
\end{figure}
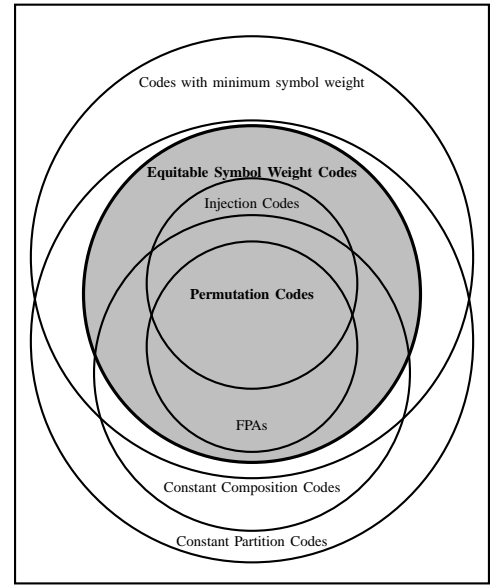

\subsection{Hamming Distance}
Consider the space $\Sigma^n$ with the distances between words measured in terms of Hamming distance.
A $q$-ary code of length $n$ and 
distance $d$ is called an $(n,d)_q$-\emph{code}, while
a $q$-ary code of length $n$ having bounded symbol weight $r$ and 
distance $d$ is called an $(n,d,r)_q$-\emph{symbol weight code},
and a $q$-ary equitable symbol weight code of length $n$
and distance $d$ is called an $(n,d)_q$-\emph{equitable symbol weight code}.

\begin{rem}
The notion of symbol equity used here differs from the notion of symbol
equity that is used in Swart and Ferriera \cite{Swart:2006}. In that work,
the authors consider the code-matrix of the code (the matrix whose rows
consist of all the codewords), and show that an equal distribution of
symbols in each column of the code-matrix results in the maximum possible
separation between all the codewords. This notion of symbol equity also
appears in the computation of the Plotkin bound on codes. In contrast, the
symbol equity discussed in this work considers the distribution of symbols
in every codeword of the code, i.e., in every row of the code-matrix.
\end{rem}

\section{Correcting Noise with MFSK Modulation}
\label{sec:MFSK}

In coded modulation for power line communications \cite{Vinck:2000}, a
$q$-ary code of length $n$ is used, whose symbols are modulated using $q$-ary MFSK.
The receiver demodulates the received signal
using an envelope detector to obtain an output, which is then decoded by a decoder.

Four detector/decoder combinations are possible: classical, modified
classical, hard-decision threshold, and soft-decision threshold
(see \cite{Ardakanietal:2010} for details). A soft-decision threshold
detector/decoder requires exact channel state knowledge and is therefore not useful if
we do not have channel state knowledge.
Henceforth, we
consider the hard-decision threshold detector/decoder here, since it contains
more information about the received signal compared to the classical and
modified classical ones. We remark that in the case of the hard-decision threshold detector/decoder,
the decoder used is a {\em minimum distance decoder}.

Let $\mathcal C$ be an $(n,d)_q$-code over alphabet $\Sigma$, and
let $\vu=(\vu_1,\dots,\vu_n)$ be a codeword transmitted over the PLC channel
where the symbol $\vu_i$ is transmitted at discrete time instance $i$ for $i\in [n]$.
The received signal (which may contain errors caused by noise) is 
demodulated to give an output 
$\vv=(\vv_1,\vv_2,\ldots, \vv_n)$ in which each $\vv_i$ is a
subset of $\Sigma$.
The errors that arise from the different types of
noise in the channel (see \cite[pp. 222--223]{Ardakanietal:2010}) have the
following effects on the output of the detector.
\begin{enumerate}
\item Narrowband noise at a particular frequency introduces a symbol at
    several consecutive discrete time instances of the transmitted signals.
The narrowband noise affects only a part of the transmission that occurs at
discrete time instances from $i=1$ to $i=n$.
Hence, narrowband noise of duration $l$ affects up to $l$ consecutive
positions in the discrete time instances from $i=1$ to $i=n$,
depending on whether the noise started prior to or during the current transmission.
Narrowband noise may be present simultaneously at multiple frequencies
corresponding to different symbols.

Let $1\le e\le q$ and $l\in \ZZ_{>0}$. If $e$ {\em narrowband noise errors of duration $l$} occur, then there
is a set $\Gamma\subseteq{\Sigma}$ consisting of $e$ symbols,
and $e$ corresponding starting instances $\{i_\sigma \le n : \sigma\in\Gamma \}$ such that for $\sigma\in\Gamma$,
\begin{equation*}
\sigma\in \vv_i \mbox{ for } \max\{1,i_\sigma\} \le i \le \min\{i_\sigma
+ l-1, n\}.
\end{equation*}

\item A signal fading error results in the absence of a symbol in the received signal. 
 Let $1\leq e\leq q$. If $e$ {\em signal fading errors} occur, then there are $e$
symbols, none of which appears in any $\vv_i$, that is,
$(\cup_{i=1}^n\vv_i)\cap\Gamma=\varnothing$ for some
$\Gamma\subseteq{\Sigma},\ |\Gamma| = e$.

\item Impulse noise results in the entire set of symbols being received at
    a certain discrete time instance.
Let $1\leq e\leq n$. If $e$ {\em impulse noise errors} occur, 
then there is a set $\Pi\subseteq{[n]}$ consisting
of $e$ positions such that $\vv_i=\Sigma$ for all $i\in\Pi$.

\item An insertion error results in an unwanted symbol in the received signal.
 Let $1\leq e\leq n(q-1)$. If $e$ {\em insertion errors} occur, then there is a set
$\Omega\subseteq{[n]\times\Sigma}$ of size $e$ such that for each $(i,\sigma)\in\Omega$, $\vv_i$ contains
$\sigma$ and $\sigma\not=\vu_i$.

\item A deletion error results in the absence of a transmitted symbol in the received signal.
Let $1\leq e\leq n$. If $e$ {\em deletion errors} occur, then there is a set
$\Pi\subseteq{[n]}$ consisting of $e$ positions such that $\vv_i$ does not contain $\vu_i$ for all
$i\in\Pi$.
\end{enumerate}
Both insertion and deletion errors are due to {\em background noise}. 
This definition of insertion and deletion error is
different from the errors that arise in an ``insertion-deletion channel''
\cite{Levenshtein:1965a}.

\begin{exa}
\begin{enumerate} Suppose $\vu=(1,2,3,4)$.
\item  Narrowband noise can start prior to or during the transmission of $\vu$.
Narrowband noise error of duration $4$ at symbol $1$ starting at discrete time instance $i=-1$
results in detector output $\vv=(\{1\}, \{1,2\}, \{3\}, \{4\})$, 
while the same narrowband noise error starting 
at discrete time instance $i=3$
results in detector output $\vv=(\{1\}, \{2\}, \{1,3\}, \{1,4\})$.
\item The same detector output can arise from different combinations of error types. 
 A signal fading error of symbol $1$ and a deletion error at position $1$
would each result in the same detector output of $\vv=(\varnothing,\{2\},\{3\},\{4\})$.
\end{enumerate}
\end{exa}
\vskip 5pt

Recall that $2^\Sigma$ denotes the power set of $\Sigma.$
For a codeword $\vu\in\Sigma^n$ and an output $\vv\in\left(2^\Sigma\right)^n$, 
define
\begin{equation*}
	d(\vu,\vv)=|\{i: \vu_i\notin \vv_i\}|.
\end{equation*}
Note that in this context, we identify $\vc\in\Sigma^n$ with 
$(\{\vc_1\},\{\vc_2\},\ldots,\{\vc_n\}) \in\left(2^\Sigma\right)^n$, 
so that $d(\vu,\vc)$ gives the Hamming distance between $\vu$ and $\vc$.
We also extend the definition of distance so that for ${\mathcal C}\subseteq \Sigma^n$,
we have $d({\mathcal C},\vv)=\min_{\vu\in{\mathcal C}}d(\vu,\vv)$.
Given $\vv\in(2^\Sigma)^n$, a minimum distance decoder (for a code $\mathcal C$)
outputs a codeword $\vu\in{\mathcal C}$ which has the smallest distance to $\vv$,
that is, a minimum distance decoder returns an element of 
\begin{equation}\label{eq:mindist}
\underset{\vu\in{\mathcal C}}{\arg \min} \ d(\vu,\vv) :=
\{\vu\in{\mathcal C}: d(\vu,\vv)\leq d(\vu',\vv) \ \forall \vu'\in{\mathcal C}\}.
\end{equation}
In the following, we study the conditions under which a minimum distance decoder
outputs the correct codeword, that is, when 
$\underset{\vu'\in{\mathcal C}}{\arg \min} \ d(\vu',\vv)=\{\vu\}$. This is equivalent
to saying that the decoder correctly outputs $\vu$ if and only if
$d({\mathcal C}\setminus\{\vu\},\vv)>d(\vu,\vv)$.

Let $d'=d({\mathcal C}\setminus\{\vu\},\vu)$.
Since $\mathcal C$ has distance
$d$, we have $d'\geq d$. Observe the following:

\begin{itemize}
\item Let $1\leq e\leq n$. If $e$ impulse noise errors occur, then in $e$
coordinates all the symbols occur. Therefore, those $e$ coordinates do not
contribute to the distance between $\vv$ and any codeword. Hence, we get
\begin{equation*}
d(\vu,\vv)=0 \quad \text{and} \quad d({\mathcal C}\setminus\{\vu\}, \vv)\geq d'-e.
\end{equation*}

\item Let $1\leq e\leq n(q-1)$. If $e$ insertion errors occur, then
there are at most $e$ coordinates which do not contribute to the distance
between $\vv$ and some codeword in the code. Hence, we get
\begin{equation*}
d(\vu,\vv)=0 \quad \text{and} \quad d({\mathcal C}\setminus\{\vu\}, \vv)\geq d'-e.
\end{equation*}

\item Let $1\leq e\leq n$. If $e$ deletion errors occur, then there are
exactly $e$ coordinates where the transmitted codeword $\vu$ differs from
$\vv$. Any other codeword still differs from $\vv$ in at least $d'$
coordinates. Therefore, we get
\begin{equation*}
d(\vu,\vv)= e \quad \text{and} \quad d({\mathcal C}\setminus\{\vu\}, \vv)\ge d'.
\end{equation*}
\end{itemize}

For errors due to narrowband noise we introduce a quantity that measures
how many coordinates of any codeword in the code are affected by the noise.
Specifically, a narrowband noise at the frequency corresponding to symbol
$\sigma$ can affect up to $n$ coordinates in a codeword, depending on the
number of times the symbol $\sigma$ appears in the codeword. If narrowband
noise is present in the set of symbols $\Gamma\subseteq\Sigma$, then the maximum
number of entries of any codeword $\vc$ that can be affected by the noise is
$\sum_{\sigma\in\Gamma} w_{\sigma}(\vc)$. Therefore, we define
\begin{equation}
\label{eq:Eec}
E(e;\C) \triangleq \max_{\vc\in\C,\;\Gamma\subseteq{\Sigma},\;|\Gamma|=e} 
            \sum_{\sigma\in\Gamma} w_\sigma(\vc).
\end{equation}
The expression $E(e;\C)$ measures the maximum number of coordinates, over
all codewords in $\C$ that are affected by $e$ narrowband noise.
Equation \eqref{eq:Eec} assumes that the duration of the narrowband noise
is at least $n$ and that it is present in all the coordinates of the codeword
transmitted.
In general, narrowband noise of duration $l$ at symbol $\sigma$ may not
be present for the full duration of the codeword. In Appendix
\ref{sec:Appendix} we show that it suffices to consider narrowband noise of
duration $n$ since it measures the maximum effect of narrowband noise on
the codewords.

Recall that $d' =d({\mathcal C}\setminus\{\vu\},\vu).$
From the definition of $E(e;\C)$, it is clear that the distance between any
codeword, other than the transmitted codeword $\vu$, and the output $\vv$
decreases by $E(e;\C)$. Similarly, in the presence of a fading error the
distance between $\vu$ and $\vv$ increases by at most $E(e;\C)$. Therefore we get the two conditions mentioned
below.
\begin{itemize}
\item Let $1\leq e\leq q$. If $e$ narrowband noise errors 
    occur, then
\begin{equation*}
d(\vu,\vv)=0 \quad \text{and} \quad d({\mathcal C}\setminus\{\vu\}, \vv)\geq d'-E(e;\C).
\end{equation*}
\item Let $1\leq e\leq q$. If $e$ signal fading errors occur, then
\begin{equation*}
d(\vu,\vv)\leq E(e;\C) \quad \text{and} \quad d({\mathcal C}\setminus\{\vu\}, \vv)\ge d'.
\end{equation*}
\end{itemize}
Hence, if we denote by $\enarrow$, $\efade$, $\eimpulse$, $\einsert$, and $\edelete$
the number of errors due to narrowband noise, signal fading, impulse noise, insertion, and deletion,
respectively, we have
\begin{align*}
d(\vu,\vv)	&\leq \edelete+E(\efade;\C),\\
d({\mathcal C}\setminus\{\vu\}, \vv)	&\geq 	d'-\eimpulse-\einsert-E(\enarrow;\C).
\end{align*}
Now,
\begin{align}
\label{ineqd}
&d(\vu,\vv)-d({\mathcal C}\setminus\{\vu\}, \vv) \notag \\ 
&\quad\le (\edelete+E(\efade;\C)) - (d'-\eimpulse-\einsert-E(\enarrow;\C)) \notag \\
 &\quad= \edelete+\eimpulse+\einsert+E(\efade;\C)+E(\enarrow;\C) - d'.
 \end{align}
Under the condition
\begin{equation*}
\edelete+\eimpulse+\einsert+E(\efade;\C)+E(\enarrow;\C) <d,
\end{equation*}
the inequality (\ref{ineqd}) reduces to $d(\vu,\vv)<d({\mathcal C}\setminus\{\vu\},\vv)$, which
implies correct decoding.

On the other hand, if 
\begin{equation*}
\edelete+\eimpulse+\einsert+E(\efade;\C)+E(\enarrow;\C)\geq d,
\end{equation*}
say $\eimpulse=d$,
and $\vu,\vw\in{\mathcal C}$ is such that $d(\vu,\vw)=d$ (since
${\mathcal C}$ has distance $d$, $\vu,\vw$ must exist), then $d'=d({\mathcal C}\setminus\{\vu\},\vu)=d$,
and we have $d(\vu,\vv)-d({\mathcal C}\setminus\{\vu\},\vv)\leq d-d'= 0$. In this case,
the correctness of the decoder output cannot be guaranteed. We therefore have the
following theorem.

\begin{thm}\label{thm:ecc}
Let $\mathcal C$ be an $(n,d)_q$-code over alphabet $\Sigma$. Let $\edelete,
\eimpulse,\einsert\in [n]$, $\enarrow,\efade \in [q]$. Then
$\mathcal C$ is able to correct $\enarrow$ narrowband noise errors, 
$\efade$ signal fading errors, 
$\eimpulse$ impulse
noise errors, $\einsert$ insertion errors, and $\edelete$ deletion errors if and
only if
\begin{equation*}
\edelete+\eimpulse+\einsert+E(\efade;\C)+E(\enarrow;\C) < d.
\end{equation*}
\end{thm}

Therefore, the parameters $n$, $q$, $d$, and $r$ (symbol weight) of a code
are insufficient to characterize the total error-correcting capability of a code in a PLC system using MFSK,
since $E(e;\C)$ cannot be specified by $n$, $q$, $d$, and $r$ alone.
We now introduce an additional new parameter that together with $n$, $q$, and $d$,
more precisely captures the error-correcting
capability of a code for PLC using MFSK.

\begin{defn} 
Let $\mathcal C$ be a code of distance $d$. 
The {\em narrowband noise error-correcting capability} of $\mathcal C$ is
\begin{equation*}
c({\mathcal C})=\min\{e: E(e;\C)\ge d\}.
\end{equation*}
\end{defn}

From Theorem \ref{thm:ecc} we infer that a code $\mathcal C$ 
can correct up to $c({\mathcal C})-1$ 
narrowband noise errors.
In general, %
the minimum value of $c(\C)$ is about $d/r$ if all the symbols
occur exactly $r$ times, and the maximum  value of $c(\C)$ is at most $d$
if all the symbols appear once. Therefore, for a code $\mathcal C$ with bounded symbol weight $r$,
we have
$\lceil{d/r}\rceil\leq c({\mathcal C}) \leq \min\{d,q\}$. However, the gap between the upper and lower
bounds can be large. Furthermore, the lower bound can be attained, giving
codes of low resilience against narrowband noise, as is shown in the following example.

\begin{exa}
The code
\begin{equation*}
\mathcal C=\{(\underbrace{1,\ldots ,1}_{r\text{ times}}, 2,3,4,\ldots, q),
(\underbrace{2,\ldots ,2}_{r\text{ times}}, 1,3,4,\ldots, q)\}
\end{equation*}
is a $(q+r-1,r+1,r)_q$-symbol weight code with narrowband noise
error-correcting capability
$c({\mathcal C})=\lceil d/r\rceil=2$. 
\end{exa}

In the rest of the paper, we write $E(\C)$ when we want to consider
$E(e;\C)$ as
a function of $e$, $E(\C):[q]\to[n]$, for a specific code $\C$.
In the next section, we provide a tight upper bound for $c({\mathcal C})$ and 
demonstrate that equitable symbol weight codes attain this upper bound.

\section{$E(\C)$ and Equitable Symbol Weight Codes}

In general, for a PLC system, narrowband noise may occur with different durations.
However, because of the result in Lemma~\ref{lem:ec}, in the rest of this section we consider only narrowband noise of duration $n$ for analysis.
In the rest of the section, we then demonstrate the optimality of equitable symbol weight codes with respect to parameter $E(\C)$.

\subsection{Relation with Symbol Weight and Partition}

Symbol weight provides an estimate for $E(\C)$.
Specifically, if $\mathcal C$ is a code of length $n$ with bounded symbol weight $r$, then 
$E(1;\C)=r$, and for $e>1$ the minimum value possible is $r+e-1$ if any other symbol
occurs exactly once. Therefore,
$E(e;\C)\geq \min\{n,r+e-1\}$.

On the other hand, if $\mathcal C$ is a constant partition code with partition $\langle c_\sigma: \sigma\in \Sigma\rangle$, 
$E(\C)$ can be determined precisely. Assume $\Sigma=[q]$ and $c_1\ge c_2
\ge \cdots \ge c_q$, then $E(e;\C)$ is the sum of $e$ largest symbol
weights in any codeword, i.e.,
\begin{equation*}
E(e;\C)=\sum_{i=1}^e c_i \mbox{ for all $e\in [q]$}. 
\end{equation*}

Further, suppose that $\C$ is an equitable symbol weight code. Then from 
Lemma \ref{lem:comp}, $\C$ has constant partition $\langle r^{q-t}(r-1)^t \rangle$, where $r=\lceil{n/q}\rceil$ and $t=qr-n$.
Hence,
\begin{equation*}
E(e;\C)=
\begin{cases}
re, 						& \text{if $e\le q-t$,}\\
r(q-t)+(e-q+t)(r-1), 	& \text{if $q-t<e\le q$.}
\end{cases}
\end{equation*}

\subsection{Importance of Symbol Equity}

For $c({\mathcal C})$ to be large, $\ec$ must grow slowly as a function of $e$.
We seek codes $\mathcal C$ for which $\ec$ grows as slowly as possible.
In this subsection we show that the minimum growth of $E(\C)$
is achieved when the
maximum symbol weight in any codeword of the code is at most $\ceiling{n/q},$
i.e., the symbols are equitably distributed in any codeword.
Fix $n,q$, and let $\mathcal F_{n,q}$ be the (finite) family of functions
\begin{equation*}
\mathcal F_{n,q}=\{E(\C): \mbox{$\mathcal C$ is a $q$-ary code of length $n$} \}.
\end{equation*}
If $f\in \mathcal F_{n,q}$, then $f$ is a monotone increasing function
with $f(q)=n$. 
We say that $f\prec g$ if 
\begin{multline}\label{eq:order}
\mbox {there exists
$e'\in[q]$ with  $f(e)=g(e)$ for $e\leq e'-1$, }\\
\mbox{and $f(e')<g(e')$.}
\end{multline}
Define the total order $\preceq$ on ${\mathcal F}_{n,q}$ so that
$f\preceq g$ if either $f(e)=g(e)$ for all $e\in[q]$ or $f\prec g$.

The following proposition states that the total order $\preceq$, in some sense, orders codes of same length and alphabet size
in accordance to their capabilities in a PLC system.

\begin{prop}\label{prop:order}
Let $\C$ and $\C'$ be $(n,d)_q$-codes. Suppose $E(\C)\prec E(\C')$ with $e'$ satisfying equation (\ref{eq:order}). If $E(e';\C)< d$,
then there exists a set of errors that 
$\C$ is able to correct but $\C'$ is unable to correct.
\end{prop}

\begin{IEEEproof}
Consider $e'$ narrowband noise errors of duration $n$ and $d-E(e';\C)-1$ impulse errors.
Then $E(e';\C)+(d-E(e';\C)-1)<d$, but $E(e';\C')+(d-E(e';\C)-1)\ge d$. 
The proposition then follows from Theorem \ref{thm:ecc}.
\end{IEEEproof}

Hence we seek the least element
in ${\mathcal F}_{n,q}$ with respect to the total order $\preceq$.

\begin{prop}\label{prop:fmin}
Let $f^*_{n,q}:[q]\rightarrow[n]$ be defined by
\begin{equation*}
\small
f^*_{n,q}(e)=
\begin{cases}
re,&\text{if $1\leq e\leq q-t$,} \\
r(q-t)+(e-q+t)(r-1),&\text{otherwise},
\end{cases}
\end{equation*}
where $r=\lceil{n/q}\rceil$ and $t=qr-n$. Then $f^*_{n,q}$ is the unique least element
in ${\mathcal F}_{n,q}$ with respect to the total order $\preceq$.
\end{prop}

\begin{IEEEproof}
Since $\preceq$ is total, it suffices to establish that
$f^*_{n,q}\preceq f$ for all $f\in{\mathcal F}_{n,q}$, and that $f^*_{n,q}\in{\mathcal F}_{n,q}$.

Let $f=\ec\in{\mathcal F}_{n,q}$, 
where $\mathcal C$ is a $q$-ary code of length $n$ over the alphabet $[q]$.
Let $\vu\in{\mathcal C}$. By permuting symbols if necessary, we may assume that
$w_1(\vu) \geq w_2(\vu) \geq\cdots\geq w_q(\vu)$. We show that for all $e\in[q]$,
\begin{equation}
\label{ineq:esw}
\sum_{i=1}^e w_i(\vu)\geq f^*_{n,q}(e). 
\end{equation}

Suppose on the contrary that $\sum_{i=1}^e w_i(\vu) < f^*_{n,q}(e)$ for some $e\in[q]$.
If $e\leq q-t$, then we have $\sum_{i=1}^e w_i(\vu)<re$ and 
$r-1\geq w_e(\vu)\geq w_{j}(\vu)$ for $j\geq e+1$. Hence,
\begin{equation*}
n = \sum_{i=1}^q w_i(\vu) < re+(q-e)(r-1)=qr-q+e\leq qr-t=n,
\end{equation*}
a contradiction.

Similarly, when $e> q-t$, we have $\sum_{i=1}^e w_i(\vu)<r(q-t)+(e-q+t)(r-1)$ 
and $r-1\geq w_e(\vu)\geq w_{j}(\vu)$ for $j\geq e+1$. Hence,
\begin{align*}
n&= \sum_{i=1}^q w_i(\vu)\\
&< 	r(q-t)+(e-q+t)(r-1)+ (q-e)(r-1)\\
&=		qr-t=n,
\end{align*} 
also a contradiction. Hence, \eqref{ineq:esw} holds. This then implies $E(e;\C)\geq f^*_{n,q}(e)$ for all $e\in[q]$, 
and consequently $f\succeq f^*_{n,q}$.

The proposition then follows by noting that $f^*_{n,q}\in \mathcal F_{n,q}$, since
$\ec=f^*_{n,q}$ when $\mathcal C$ is a $q$-ary equitable symbol weight code of length $n$.
\end{IEEEproof}
\vskip 5pt

\begin{cor}\label{cor:esw}
$\mathcal C$ is a $q$-ary equitable symbol weight code of length $n$ if and only if $\ec=f^*_{n,q}$.
\end{cor}

\begin{IEEEproof}
If $\mathcal C$ is a $q$-ary equitable symbol weight code of length $n$, we have
already determined that $\ec=f^*_{n,q}$.
Hence, it only remains to show that $\ec=f^*_{n,q}$ 
implies $\mathcal C$ is a $q$-ary equitable symbol weight code of length $n$. 
Let $\vu\in \mathcal C$ and we follow the notation in the proof of Proposition \ref{prop:fmin}. 
Equality holds in \eqref{ineq:esw} if and only if $w_i(\vu)=r$ 
for $1\leq i\leq q-t$ and $w_i(\vu)=r-1$, otherwise. That is, $\vu$ has equitable symbol weight. 
Hence, $\mathcal C$ is an equitable symbol weight code.
\end{IEEEproof}

It follows that an equitable symbol weight code $\mathcal C$ gives $\ec$
of  the slowest growth
rate. From Proposition \ref{prop:order}, this is the desired condition for correcting as many
narrowband noise and signal fading errors as possible.

We end this section with a tight upper bound on $c({\mathcal C})$.

\begin{cor}\label{cor:esw2}
Let $\mathcal C$ be an $(n,d)_q$-code. Then
\begin{equation*}
c({\mathcal C})\leq \min\,\{e: f^*_{n,q}(e)\geq d\},
\end{equation*}
and equality is achieved when $\mathcal C$ is an equitable symbol weight code.
\end{cor}

\begin{IEEEproof}
Let $c' =\min\{e: f^*_{n,q}(e)\geq d\}$. 
Observe that 
\begin{equation*}
E(c';\C) \geq f^*_{n,q}(c')\geq d.
\end{equation*}
Hence, by minimality of $c({\mathcal C})$, we have $c({\mathcal C})\leq c'$. 
The second part of the statement follows from Corollary \ref{cor:esw}.
\end{IEEEproof}

The results in this section establish that an equitable
symbol weight code has the best narrowband noise and signal fading error-correcting capability,
among codes of the same distance and symbol weight.

\section{Simulation Results}

In this section, we study the performance of equitable symbol weight codes
in a simulated setup.
The setup is as follows. We transmit with a code of length $n$ over alphabet $\Sigma$.
Let $p$ be a real number between $0$ and $1$ and $L=\{b n: b\in [10]\}$. We simulate a PLC channel with the following 
characteristics:
\begin{enumerate}
\item for each $\sigma \in \Sigma$, narrowband noise error\footnote{The choice of $L$ is similar to that of the narrowband noise model in the setup of  Verfeld {\em et al.}\cite{Versfeldetal:2005,Versfeldetal:2010}.} of duration $l\in L$ occurs at symbol $\sigma$ with probability $p$,
\item for each $\sigma \in \Sigma$, a signal fading error occurs at symbol
    $\sigma$ with probability $Q$,
\item for each $i \in [n]$, an impulse noise error occurs at coordinate $i$
    with probability $Q$, and 
\item for each $(\sigma,i) \in \Sigma\times [n]$, an insertion/deletion
    error occurs at symbol $\sigma$ and coordinate $i$ with probability $Q$.
\end{enumerate}
These errors occur independently.

We choose $10^5$ random codewords (with repetition) 
from each code to transmit through
the simulated PLC channel. At the receiver, we decode the detector output $\vv$ to the
codeword $\vu'$ using the minimum distance decoder defined in equation (\ref{eq:mindist}). 
The number of symbols in error is then $d(\vu',\vu)$ and
the \emph{symbol error rate} is the ratio of the total number of symbols in
error to the total number of symbols transmitted. 

{\em Decoding with narrowband noise detection}: Versfeld \etal \cite{Versfeldetal:2005,Versfeldetal:2010}
introduced a method to detect narrowband noise in order to enhance the error correction capability of the detector introduced in Section \ref{sec:MFSK}, when used with bounded distance decoding.
Based on the energy metrics obtained at each time slot for each frequency, they first determine the presence of narrowband interference
and if so, the metrics of the corresponding frequency are set to zero.
Depending on the detector/decoder combination, a signal is sent to the decoder.
Specifically, consider narrowband noise detection with the use of an $(n,d,r)$-symbol weight code. 
If the number of discrete time instances in which a particular symbol appears, exceeds $\floor{(n+r)/2}$,
 the particular symbol is removed from the coordinates in which it occurs.
We describe an algorithm
 to detect and remove narrowband noise in Algorithm \ref{fig:nbdetect}.

\begin{algorithm}[!h] 
\SetAlgoLined
\KwIn{Detector Output, $\vv\in(2^\Sigma)^n$} 
\KwOut{Modified $\vv\in(2^\Sigma)^n$}
\BlankLine
$\tau\gets \floor{(n+r)/2}$\; \For{$\sigma\in\Sigma$}{
\If{$|\{i:\sigma\in \vv_i\}|>\tau$}{\For{$i\in[n]$}{$\vv_i\gets \vv_i\setminus\{\sigma\}$}}} 
\caption{Narrowband noise detection with an $(n,d,r)$-symbol weight code}
\label{fig:nbdetect}
\end{algorithm}

\subsection{Minimum Symbol Weight Codes}

We exhibit the difference in performance between 
equitable symbol weight codes and (non-equitable) minimum symbol weight codes.
Specifically, we consider the codes of various lengths and relative distances in Table \ref{tab:msw}.

\begin{table*}[!t]
\centering
\caption{Comparison of Equitable Symbol Weight Codes and Minimum Symbol Weight Codes}
\footnotesize
\label{tab:msw}
\begin{tabular}{|c|C{0.8cm}|C{1cm}|C{3cm}|C{1cm}|C{1.5cm}|C{0.8cm}|c|}
\hline
Code & Length & Distance & Narrowband noise error-correcting capability & Symbol weight & Alphabet size & Size & Remarks\\ 
\hline
ESW$(25,24,2)_{17}$ & 25 & 24 & 16 & 2 & 17 & 51 & equitable symbol weight\\
MSW$(25,24,2)_{17}$ & 25 & 24 & 12 & 2 & 17 & 51 & minimum symbol weight\\ \hline
ESW$(11,6,2)_{10}$ & 11 & 6 & 5 & 2 & 10 & 1000 & equitable symbol weight\\
MSW$(11,6,2)_{10}$ & 11 & 6 & 3 & 2 & 10 & 1000 & minimum symbol weight\\ \hline
\end{tabular}
\end{table*}

\begin{figure}[!h]
    \centering
    \scalebox{0.43}{\includegraphics[keepaspectratio]{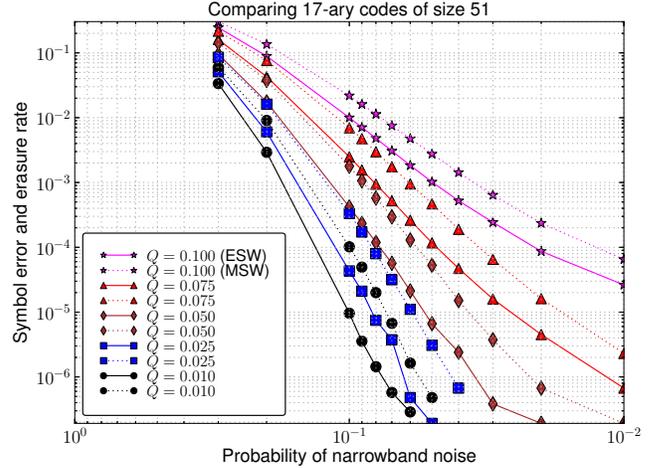}}
    \caption{Comparison of equitable symbol weight codes (solid lines) and
        minimum symbol weight codes (dashed lines) with varying
    probabilities of noise}
    \label{fig:comparison-prob}
\end{figure}

In Fig.~\ref{fig:comparison-prob} we show the difference between the
performance of the codes
for varying probability of narrowband noise. The different plots correspond
to the probability of background noise, impulse noise and fading fixed at
$Q \in \{0.1, 0.075, 0.05, 0.025, 0.01\}$. The solid lines correspond to
equitable symbol weight codes and the dotted lines correspond to minimum
symbol weight codes. Only for this particular simulation $10^7$ codewords
are transmitted.  In the simulations we detect the presence of narrowband noise\footnote{As discussed in Section \ref{sec:MFSK}, after narrowband noise detection, 
the multivalued output is given directly to a minimum distance decoder. This deviation from the setup by Versfeld \etal\ 
(where envelope detection and Viterbi threshold ratio test is applied prior to decoding) 
means that the results are independent of the choice of demodulation rule.}
using Algorithm~\ref{fig:nbdetect}.

For the rest of the simulations we fix $Q=0.05$. For equitable and minimum symbol weight codes of size $1000$, the results
of the simulation are displayed in Fig.~\ref{fig:comparison}.
The solid lines correspond to simulations in which we detect narrowband
noise and are labelled by ``(NB)''.
The dashed lines denote simulations without narrowband noise detection.
From the results, observe that
ESW$(25,24,2)_{17}$ and ESW$(11,6,2)_{10}$ 
achieve lower symbol error rates
compared to MSW$(25,24,2)_{17}$ and MSW$(11,6,2)_{10}$, respectively.

\subsection{Cosets and Subcodes of Reed-Solomon Codes}

Versfeld \etal \cite{Versfeldetal:2005,Versfeldetal:2010} showed empirically that 
using narrowband detection,
low symbol weight cosets of Reed-Solomon codes 
outperform normal Reed-Solomon codes
in the presence of narrowband noise and additive white Gaussian noise.
We continue this investigation and observe the difference in performance 
between equitable symbol weight codes and  
low symbol weight cosets of Reed-Solomon codes.
In addition, we consider subcodes of Reed-Solomon codes with low symbol weight.
In all these simulations we fix $Q=0.05,$ and vary the probability $p$ of
narrowband noise.

\begin{table*}[!ht]
\centering
\footnotesize
\caption{Comparison of Equitable Symbol Weight Codes and Low Symbol Weight Cosets and Subcodes of Reed-Solomon Codes}
\label{tab:rs}
\begin{tabular}{|c|C{0.8cm}|C{1cm}|C{3cm}|C{1cm}|C{1.5cm}|C{0.8cm}|c|}
\hline
Code & Length & Distance & Narrowband noise error-correcting capability & Symbol weight & Alphabet size & Size & Remarks\\ \hline
ESW$(7,5,1)_8$ & 7 & 5 & 5 & 1 & 8 & 336  & equitable symbol weight\\
RSC$(7,6,2)_8$ & 7 & 6 & 3 & 2 & 8 & 64 & coset of Reed-Solomon code\\
RSS$(7,5,2)_8$ & 7 & 5 & 3 & 2 & 8 & 336 & subcode of Reed-Solomon code\\ \hline

ESW$(7,2,1)_8$ & 7 & 2 & 2 & 1 & 8 & 20160 & equitable symbol weight\\
RSC$(7,4,4)_8$ & 7 & 4 & 1 & 4 & 8 & 4096 & coset of Reed-Solomon code\\
RSS$(7,3,2)_8$ & 7 & 3 & 2 & 2 & 8 & 20160 & subcode of Reed-Solomon code\\ \hline

ESW$(15,11,1)_{16}$ & 15 & 11 & 11 & 1 & 16 & 21120 & equitable symbol weight\\
RSC$(15,13,3)_{16}$ & 15 & 13 & 5  & 3 & 16 & 4096 & coset of Reed-Solomon code\\
RSS$(15,12,3)_{16}$ & 15 & 12 & 4 & 3 & 16 & 21120 & subcode of Reed-Solomon code\\ \hline
\end{tabular}
\end{table*}

Specifically, we consider the codes in Table \ref{tab:rs}. 
See \cite{Versfeldetal:2005,Versfeldetal:2010} for the construction of Reed-Solomon coset codes, denoted by RSC.
The codes denoted by RSS are subcodes of Reed-Solomon codes. 
They are obtained by expurgation of a Reed-Solomon code and retaining only the codewords with low symbol weight.

We note that it is not possible for equitable symbol weight codes and Reed-Solomon coset codes of 
the same minimum distance and length over the same alphabet to be of the same size.
Therefore, for each Reed-Solomon coset codes, we make comparisons with an equitable symbol weight code of a larger size, 
albeit with a smaller distance. However, these equitable symbol weight codes have larger narrowband noise error-correcting capabilities.
In addition, we make comparisons with subcodes of Reed-Solomon codes with parameters 
as close as possible to the corresponding equitable symbol weight codes. 
In particular, we ensure that the subcodes and the equitable symbol weight codes have the same size.

\begin{figure}
\begin{center}
\includegraphics[scale=0.43]{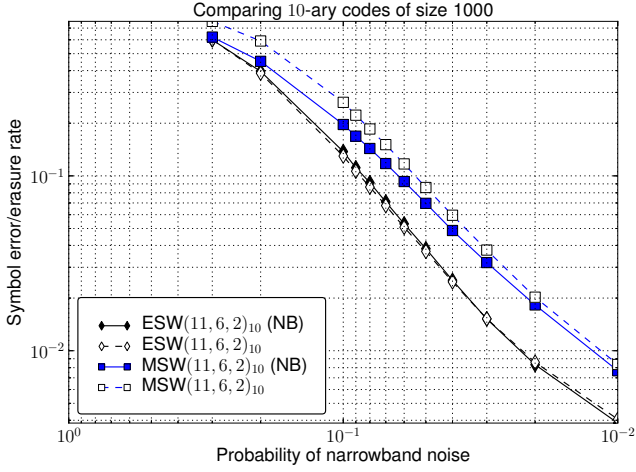}
\caption{Comparison of equitable and minimum symbol weight codes}
\label{fig:comparison}
\end{center}
\end{figure}

\begin{figure}
\begin{center}
\includegraphics[scale=0.43]{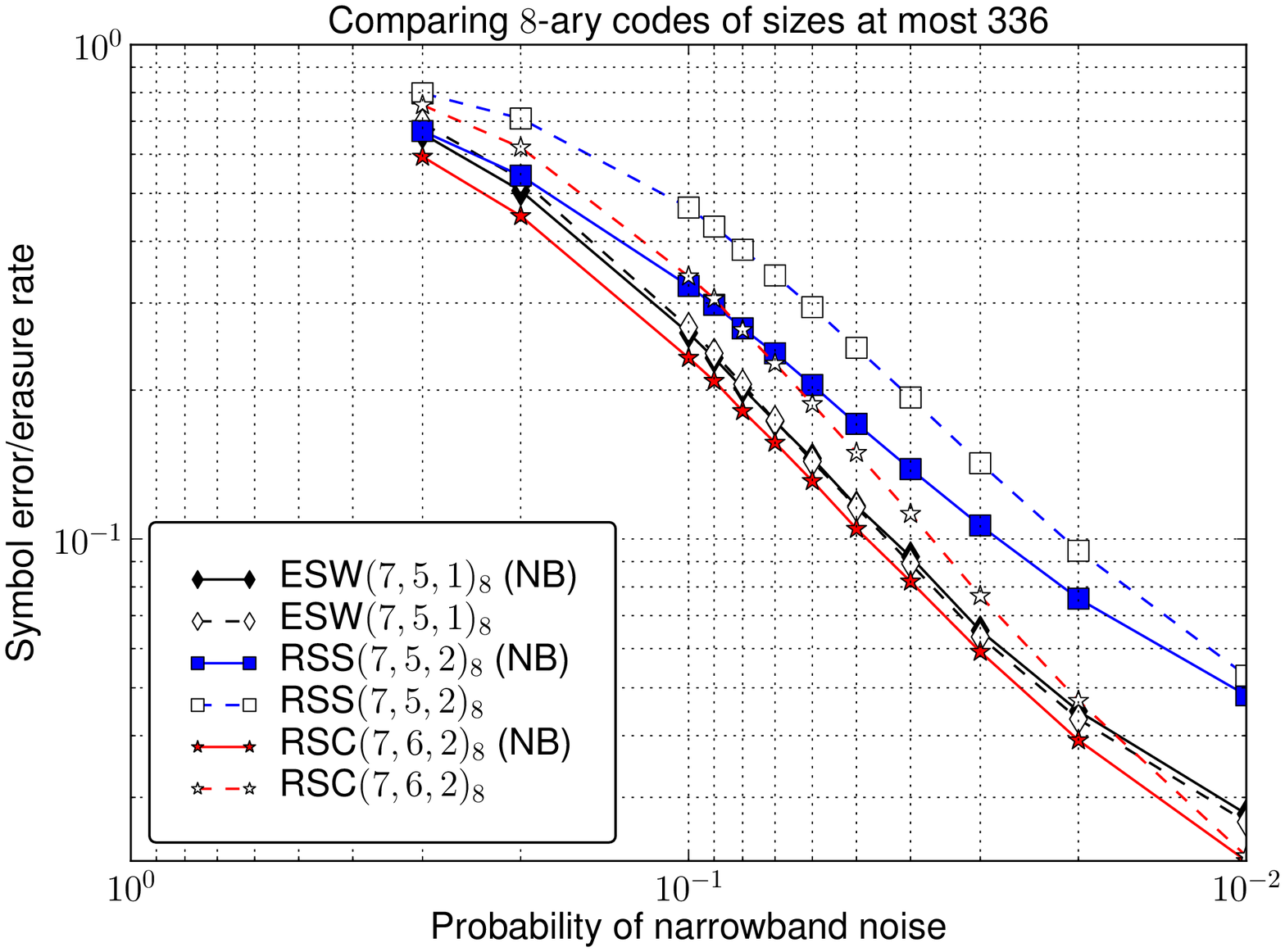}
\vspace{3pt}
\includegraphics[scale=0.43]{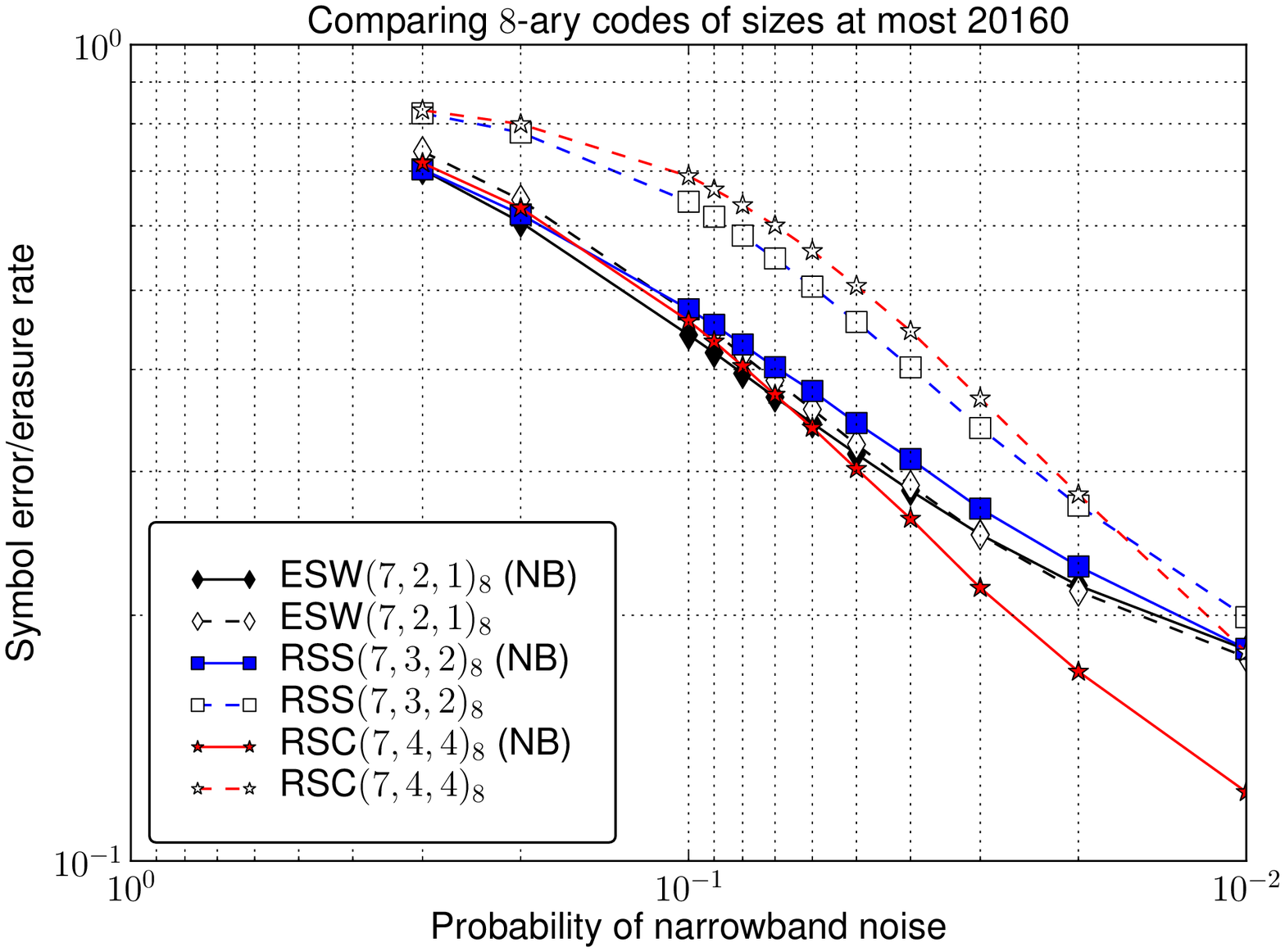}
\vspace{3pt}
\includegraphics[scale=0.43]{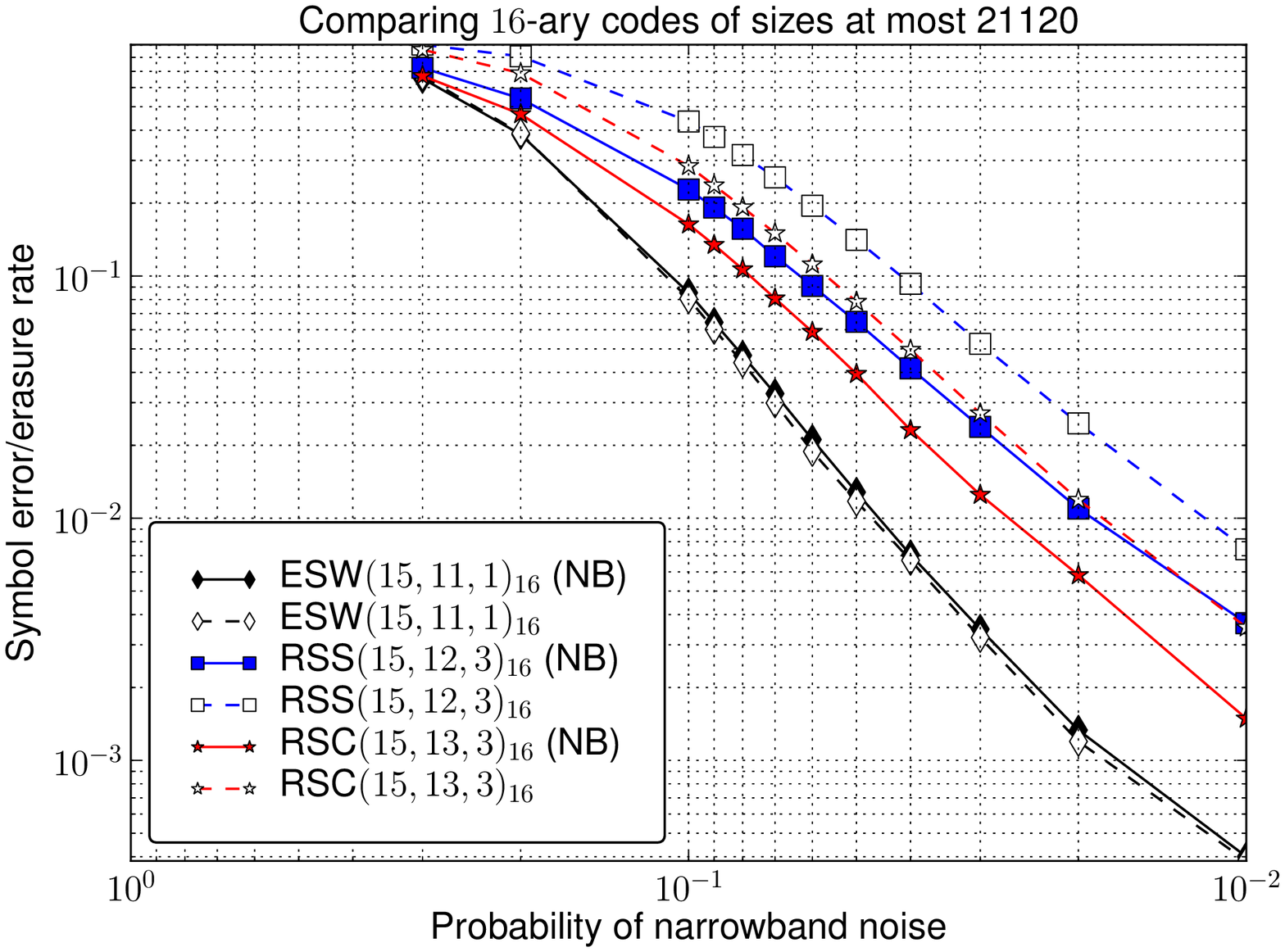}
\caption{Comparison of equitable symbol weight codes and low symbol weight cosets and subcodes of Reed-Solomon codes}
\label{fig:comparison2}
\end{center}
\end{figure}

The results of the simulation are displayed in Fig. \ref{fig:comparison2}, 
where we adopt similar conventions as in Fig. \ref{fig:comparison},
and we make the following observations.
\begin{enumerate}[(i)]
\item While narrowband noise detection in general improves the performance of codes in PLC,
it has negligible effect on the performance of equitable symbol weight codes.
A natural question is if there is another parameter that measures this improvement and 
if this parameter is related to symbol equity.
\item Equitable symbol weight codes show larger improvement over
Reed-Solomon coset codes at higher narrowband noise probabilities.
This reflects the relevance of narrowband noise error-correcting capabilities as a measure of performance
when the effects of narrowband interference are significant.
In contrast, when the effects of narrowband interference are negligible, 
the classical Hamming distance parameter provides a better measure of performance. 
\end{enumerate}

\subsection{Simulation in the presence of cyclostationary noise}
By definition, the parameter $c(\C)$ of a code $\C$ captures the
performance of the code in the presence of narrowband noise and fading. It
captures a ``worst-case error'' performance, similar to how the minimum
distance of a code determines the worst-case error performance under
bounded distance decoding. A natural question arises about how a code $\C$
with a narrowband noise error correcting capability $c(\C)$ performs in the presence of cyclostationary noise
(periodically varying noise) compared to a code $\C'$ with a lower value
of $c(\C')$. We compare the performance of the equitable symbol weight
code ESW$(25, 24, 2)_{17}$ and the minimum symbol weight code
MSW$(25, 24, 2)_{17}$ under the presence of cyclostationary noise.

\begin{figure}
\centering
\includegraphics[scale=0.43]{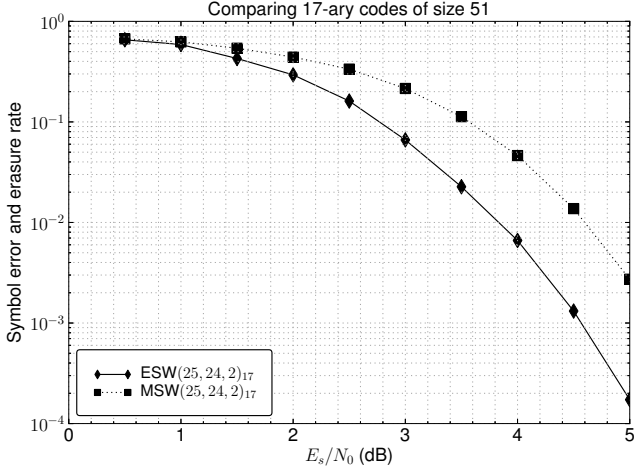}
\caption{Comparison of equitable and minimum symbol weight codes under
cyclostationary noise}
\label{fig:cyclo}
\end{figure}

The setup is as follows. A model for cyclostationary noise in the power
line channel is presented in \cite{Katayama:2006}. Gaussian noise is
generated with instantaneous variance 
\begin{multline*}
\sigma^2(t) = 0.23 + 1.38
\left|\sin\left(2\pi\frac t{T_{AC}} - 0.10\right)\right|^{1.91}\\ + 
7.17 \left|\sin\left(2\pi\frac t{T_{AC}}
- 0.61\right)\right|^{157000},
\end{multline*}
where $T_{AC} = 1/60\,\text{s}$ is the period of the mains voltage. The
instantaneous variance has a period of $T_{AC}/2,$ and has an average
variance of one, when averaged over this period. The generated Gaussian
noise is then passed through a filter with amplitude response $H(f)
= \sqrt{a/2} e^{-a|f|/2}$.,  where $a = 1.2\times 10^{-5}$.
Let the alphabet of the codes be the set $[17]$. We require $17$ individual
center frequencies to represent each symbol from the alphabet. The
transmitted signals are modulated according to the sinusoidal waves
$$
s_m(t) = \sqrt{\frac {2E_s}{T_s}}\cos(2\pi f_m t), \ t\in[0,T_s), m\in[17],
$$
where $E_s$ is the symbol energy, $T_s$ is the symbol time period, and $f_m$ are the center frequencies.
The time period of each symbol is taken to be $T_s
= 1/9 \times T_{AC}/2$, and the signal is sampled at the rate $500\times
1080\,$Hz, which is slightly above $500$ kHz. This sampling rate is
chosen to give the same integral number of samples in each $T_{AC}/2$ period.
The center frequency corresponding to the symbol $m$ is taken to be
$10.8m\,\text{kHz}$, for $m\in[17]$. This maintains a frequency separation
of $10.8\,\text{kHz}$ that is an integral multiple of $1/T_s$ and ensures
a correlation of zero between the different signal waveforms.
At each center frequency we use a square-law detector and declare one if it
detects an energy greater than $E_s/4$, otherwise it declares
a zero\footnote{This threshold is similar to the threshold used in
\cite{Vinck:2000}.}. While decoding, the codes detect the presence of
narrowband noise using Algorithm~\ref{fig:nbdetect}. The output of
the simulation is presented in Fig.~\ref{fig:cyclo}. The horizontal axis
corresponds to the signal to average noise ratio (in
dB), where the average noise power spectral density is denoted by $N_0$. We
observe that the equitable symbol weight codes outperform minimum symbol
weight codes when the noise process is cyclostationary.

\section{Conclusion}

We have introduced a new code parameter that captures the error-correcting capability 
of a code with respect to narrowband noise and signal fading. 
Equitable symbol weight codes are shown to be optimal with respect to this parameter when code length, 
alphabet size and distance are fixed. 
We also provide simulations that show equitable symbol weight codes 
achieve lower symbol error rates as compared to their non-equitable counterparts. 
These results motivate the study of equitable symbol weight codes as 
a viable option to handle narrowband noise and signal fading in a PLC channel.

\appendices
\section{Narrowband noise of different durations and $E(e;\mathcal C)$}
\label{sec:Appendix}
In this Appendix we show that it suffices to consider narrowband noise of
length $n$ instead of smaller lengths since it measures the maximum effect
of narrowband noise on the codewords. For integers $k,n,\ k\le n$, we
use the notation $[k,n] \triangleq \{k,\dots,n\}$.
Therefore, given $n$ and for an integer
$i_\sigma\le n$, we can write $\big\{i:\max\{1,i_\sigma\}\le i \le
\min\{i_\sigma+l-1,n\} \big\} = [i_\sigma, i_\sigma+l-1]\cap[n]$.
For errors due to narrowband noise, we define the following quantity for
$\Gamma\subset\Sigma$, $l\in \ZZ_{>0}$, $\vc\in\C$,
\begin{equation*}
E(\Gamma;l,\vc)=\max_{i_\sigma\le n:\,\sigma\in\Gamma}\left |\{i: i\in
   [i_\sigma,i_\sigma+l-1]\cap [n], \vc_i=\sigma\}\right|.
\end{equation*}
The quantity $E(\Gamma;l,\vc)$ measures the maximum number of
coordinates in $\vc$ that can be affected by narrowband noise of duration $l$ at symbols in $\Gamma$.

Let $L\subset \ZZ_{>0}$. We consider the following quantity as a function
in $e$, $$E(L,\C):[q]\to [n],$$
\begin{equation*}
E(e;L,\C)=\max_{\Gamma \subseteq{\Sigma},\;|\Gamma|=e,\;l\in L,\;\vc\in\C} E(\Gamma;l,\vc),
\end{equation*}
\noindent then $E(e;L,\C)$ measures the maximum number of
coordinates, over all codewords in $\mathcal C$,
that can be affected by $e$ narrowband noise of duration $l\in L$. 
The following lemma states that it suffices to consider the maximum duration when determining
the performance of a code in a PLC. 

\begin{lem}\label{lem:ec}
Let $\mathcal C$ be a $q$-ary code of length $n$. Consider $L\subset \ZZ_{>0}$ and define
$n'=\min \{n, \max L\}$. Then 
\begin{equation*}
E(L,\C)=E(\{n'\},\C).
\end{equation*}
\end{lem}

\begin{IEEEproof}
Let $l'=\max L$ and fix $l\in L$ and $e\in [q]$. 

Observe that since
$[i,i+l-1]\subseteq [i,i+l'-1]$ for $i\le n$,
\begin{equation*}
E(\Gamma;l,\vc) \le E(\Gamma;l',\vc) \mbox{ for $\vc\in \C$,
$\Gamma\subset\Sigma$}.
\end{equation*}
Hence, $E(e;\{l\},\C)\le E(e;\{l'\},\C)$ and so, $E(e;L,\C) \le E(e;\{l'\},\C)$. 

In addition, since $[i,i+l-1]\cap [n]\subseteq [n]$ for $i\le n$,
\begin{equation*}
E(\Gamma;l,\vc) \le E(\Gamma;n,\vc) \mbox{ for $\vc\in \C$,
$\Gamma\subset\Sigma$}.
\end{equation*}
Similar argument shows that $E(e;L,\C) \le E(e;\{n\},\C)$. 
Since $l'\in L$, we have $E(e;L,\C) \ge E(e;\{l'\},\C)$ and the lemma follows.
\end{IEEEproof}

The following is now immediate.

\begin{cor}\label{cor:ec}
Let $\mathcal C$ be a $q$-ary code of length $n$.
For $L\subset \ZZ_{>0}$, 
\begin{equation*}
E(e;L,\C)\le E(e;\{n\},\C) \mbox{ for all }e\in [q].
\end{equation*}
\end{cor}

Therefore, $E(L,\C)$, which measures the maximum effect of narrowband noise on codewords, is maximized
when $L=\{n\}$.
Hence, we assume that only narrowband noise of duration $n$ occurs.

\section*{Acknowledgement}
The authors thank Han Vinck for useful discussions.
The authors are also grateful to the anonymous reviewers and Editor
Prof. Stefano Galli for their constructive and insightful comments, which
helped improve the presentation of this work substantially.




\begin{IEEEbiographynophoto}{Yeow Meng Chee}
(SM'08) received the B.Math. degree in computer science and combinatorics and optimization 
and the M.Math. and Ph.D. degrees in computer science, from the University of Waterloo, Waterloo, ON, Canada, 
in 1988, 1989, and 1996, respectively.

He is currently Professor and Chair of the School of Physical and Mathematical Sciences, Nanyang Technological University, Singapore. 
Prior to this, he was Program Director of Interactive Digital Media R\&D in the Media Development Authority of Singapore, 
Postdoctoral Fellow at the University of Waterloo and 
IBM's Z\"urich Research Laboratory, General Manager of the Singapore Computer Emergency Response Team, and 
Deputy Director of Strategic Programs at the Infocomm Development Authority, Singapore. 
His research interest lies in the interplay between combinatorics and computer science/engineering, particularly combinatorial design theory, coding theory, extremal set systems, and electronic design automation.
\end{IEEEbiographynophoto}

\begin{IEEEbiographynophoto}{Han Mao Kiah}
received the B.Sc.(Hon) degree in mathematics from the National University of Singapore, Singapore in 2006. 
Currently, he is working towards his Ph.D. degree at the Division of Mathematical Sciences, School of Physical and Mathematical Sciences, Nanyang Technological University, Singapore.
His research interest lies in the application of combinatorics to engineering problems in information theory. 
In particular, his interests include combinatorial design theory, coding theory and power line communications.
\end{IEEEbiographynophoto}

\begin{IEEEbiographynophoto}{Punarbasu Purkayastha}
(M'10) received the B.Tech. degree in electrical engineering from Indian
Institute of Technology, Kanpur, India in 2004, and the Ph.D. degree in
electrical engineering from University of Maryland, College Park, U.S.A.,
in 2010.

Currently, he is a Research Fellow at the Division of Mathematical
Sciences, School of Physical and Mathematical Sciences,  Nanyang
Technological University, Singapore. His research interests include coding
theory, combinatorics, information theory and communication theory.
\end{IEEEbiographynophoto}

\begin{IEEEbiographynophoto}{Chengmin Wang} 
received the B.Math. and Ph.D. degrees in mathematics from Suzhou University, China in 2002 and 2007, respectively.
Currently, he is an Associate Professor at the School of Science, Jiangnan University, China. 
Prior to this, he was a Visiting Scholar at the School of Computing, Informatics and Decision Systems Engineering, 
Arizona State University, USA, from 2010 to 2011 and was a 
Research Fellow at the Division of Mathematical Sciences, 
School of Physical and Mathematical Sciences, Nanyang Technological University, Singapore from 2011 to 2012. 
His research interests include combinatorial design theory and its applications in coding theory and cryptography.
\end{IEEEbiographynophoto}

\end{document}